\documentclass[aps,prl,twocolumn,showpacs,superscriptaddress]{revtex4}
\usepackage{graphicx}
\usepackage{epsfig}
\usepackage{epstopdf}
\usepackage{color}
\usepackage{amsfonts}
\usepackage{amsthm}
\usepackage{amsmath}
\usepackage{amssymb}
\usepackage[latin1]{inputenc}
\usepackage[english]{babel}

\begin{document}


\def\beq{\begin{equation}}
\def\eeq{\end{equation}}
\def\be{\begin{equation}}
\def\ee{\end{equation}}

\def\iomn{i\omega_n}
\def\iom#1{i\omega_{#1}}
\def\c#1#2#3{#1_{#2 #3}}
\def\cdag#1#2#3{#1_{#2 #3}^{+}}
\def\epsk{\epsilon_{{\bf k}}}
\def\Ga{\Gamma_{\alpha}}
\def\Seff{S_{eff}}
\def\dinf{$d\rightarrow\infty\,$}
\def\T{\mbox{Tr}}
\def\t{\mbox{tr}}
\def\cG0{{\cal G}_0}
\def\cS{{\cal S}}
\def\divnum{\frac{1}{N_s}}
\def\vac{|\mbox{vac}\rangle}
\def\intR{\int_{-\infty}^{+\infty}}
\def\intb{\int_{0}^{\beta}}
\def\spinup{\uparrow}
\def\spindown{\downarrow}
\def\bra{\langle}
\def\ket{\rangle}

\def\ka{{\bf k}}
\def\vk{{\bf k}}
\def\vq{{\bf q}}
\def\vQ{{\bf Q}}
\def\vr{{\bf r}}
\def\q{{\bf q}}
\def\R{{\bf R}}
\def\kp{\bbox{k'}}
\def\a{\alpha}
\def\b{\beta}
\def\d{\delta}
\def\D{\Delta}
\def\e{\varepsilon}
\def\eps{\epsilon}
\def\ed{\epsilon_d}
\def\ef{\epsilon_f}
\def\g{\gamma}
\def\G{\Gamma}
\def\l{\lambda}
\def\L{\Lambda}
\def\o{\omega}
\def\ph{\varphi}
\def\s{\sigma}
\def\chib{\overline{\chi}}
\def\et{\widetilde{\epsilon}}
\def\hn{\hat{n}}
\def\hnu{\hat{n}_\uparrow}
\def\hnd{\hat{n}_\downarrow}

\def\hc{\mbox{h.c}}
\def\Im{\mbox{Im}}

\def\est{\varepsilon_F^*}
\def\v2o3{V$_2$O$_3$}
\def\uc2{$U_{c2}$}
\def\uc1{$U_{c1}$}


\def\bea{\begin{eqnarray}}
\def\eea{\end{eqnarray}}
\def \bal{\begin{align}}
\def \eal{\end{align}} 
\def\#{\!\!}
\def\@{\!\!\!\!}

\def\vi{{\bf i}}
\def\vj{{\bf j}}

\def\+{\dagger}


\def\up{\spinup}
\def\down{\spindown}


\def\'{\prime}
\def\"{{\prime\prime}}


\title{Orbital-Selective Mott transition out of band degeneracy lifting}
\author{Luca de' Medici}
\affiliation{Department of Physics and Center for Materials Theory, Rutgers University, Piscataway NJ 08854, USA}
\author{S.~R.~Hassan}
\affiliation{Department de Physique, Université de Sherbrooke, Québec, Canada J1K 2R1}
\author{Massimo Capone}
\affiliation{SMC, CNR-INFM, and Università di Roma "La Sapienza'',  Piazzale Aldo Moro 2, I-00185 Roma, Italy}
\affiliation{ ISC-CNR, Via dei Taurini 19, I-00185 Roma, Italy}
\author{Xi Dai}
\affiliation{Beijing National Laboratory for Condensed Matter
Physics, and Institute of Physics, Chinese Academy of Sciences,
Beijing 100190, China}

\begin{abstract}
We outline a general mechanism for  Orbital-selective Mott transition (OSMT), the coexistence of both itinerant and localized conduction electrons, and show how it can take place in a wide range of realistic situations, even for bands of
identical width and correlation, provided a crystal field splits the energy
levels in manifolds with different degeneracies and the exchange
coupling is large enough to reduce orbital fluctuations. The mechanism
relies on the different kinetic energy in manifolds with different degeneracy.
This phase has Curie-Weiss susceptibility and non Fermi-liquid behavior,
which disappear at a critical doping, all of which is reminiscent of the physics of the pnictides.
\end{abstract}

\maketitle

The localization of electrons in partially filled bands due 
to their mutual Coulomb repulsion, or Mott metal-insulator transition, 
is perhaps the most striking effect of electron-electron interaction and
it escapes a proper description in the standard one-body theory of
solids. 
Only the development of many-body methods like slave-particle
mean-fields and Dynamical Mean-Field Theory\cite{georges_RMP_dmft}
(DMFT) allowed to draw a coherent picture at least for the single-band
Hubbard model, which accounts for many properties of three-dimensional 
oxides\cite{limelette_v2o3_science}.

On the other hand, the Mott physics in multi-band systems is quite richer,  and it presents many interesting phenomena that still lack a complete understanding.
Among them, the Orbital-Selective Mott Transition (OSMT) (i.e. the possibility of localization of part of the conduction electrons while the rest remains itinerant) is the most interesting and unique feature. The OSMT
 has been evoked\cite{Anisimov_OSMT} as a possible explanation of the seemingly mixed itinerant (metallic resistance) and localized (Curie-Weiss magnetic susceptibility)  character of the conduction bands in the triplet-superconductor $Sr_{2-x}Ca_xRuO_4$. This idea has attracted considerable attention and sparked a good deal of theoretical investigation\cite{Koga_OSMT,demedici_Slave-spins,Ferrero_OSMT,Liebsch_OSMT_3}, uncovering a rich physics in which the Hund's coupling plays an important role\cite{demedici_Slave-spins,Ferrero_OSMT} and non-Fermi liquid behavior is found in the metallic component\cite{Biermann_nfl}.
Even if the occurrence of an OSMT in $Sr_{2-x}Ca_xRuO_4$ is still under debate\cite{Dai_OSMT,Liebsch_SrRuO4_negDelta,Neupane_OSMT,Balicas_fsfceCSRO,Wang_Yang_ARPES_CaSrRuO4}, it is suggestive that other anomalous superconductors like the K-BEDT organics and the iron oxypnictides\cite{kamihara} have been considered as possible realization of this physics\cite{Shorikov}, and that a k-space selectivity has been evoked even for the high-$T_c$ cuprates\cite{Venturini_MIT_cuprates,Biermann_nfl}.

Several distinct mechanisms that can lead to an OSMT have been identified. 
The original and most obvious is that the conduction electrons lie in separate non-hybridized correlated bands of different bandwidths\cite{Anisimov_OSMT}, or that bands of similar bandwidth have different intraband Coulomb repulsion\cite{Wu_FeAs_OSMT}.
In Ref. \cite{Werner_Hund} the crystal-field splitting of two bands of equal bandwidth has been identified as a source of the OSMT. By lightly doping the Mott insulator, electrons populate the lower band, while the higher one will remain half-filled and insulating, leading to an OSMT driven by doping at incommensurate fillings. Unfortunately all these situations, in which the OSMT arises, are quite specific.

In this paper we identify and study a new mechanism for the OSMT which brings it from a rare to a much more plausible phenomenon. We
consider a system in which a crystal field splitting divides the
original degenerate manifold in subsets with different degeneracy, the
simplest examples being three levels split so that one level has
different
energy with respect to the other two, that remain degenerate (or nearly degenerate). In this
case an OSMT can take place even in the case of equal bandwidth and Coulomb repulsion for all bands, and for commensurate fillings. 
This phenomenon can be understood on the basis our knowledge about Mott
transitions for degenerate bands. 
The critical interaction strength $U_c$ for the Mott transition is indeed
larger for manifolds of bands with larger degeneracy due to their increased kinetic
energy\cite{gunnarsson_fullerenes,florens_multiorb}. For example, in the $SU(N)$-orbital Hubbard model $U_c$ scales with $N$ at large $N$, while, for fixed number of bands, a Mott transition occurs at any integer filling\cite{Lu_gutz_multiorb,rozenberg_multiorb} and $U_c$  is largest at half-filling and decreases moving away from it.

Therefore if the split manifolds were composed by different number of levels and completely decoupled they would
definitely undergo Mott transitions at distinct values of $U$.
 Here we show how this survives when the manifolds are coupled by an
 interband coulomb repulsion and a spin-spin exchange interaction, 
identifying the role of the latter and the of the crystal field to suppress orbital fluctuations, leading to an effective orbital decoupling.


The situation we consider is indeed very general, making our OSMT probably the most common in nature. It is well known that, e.g., a cubic crystal field splits the five d-obitals in two groups, $t_{2g}$ and $e_g$, respectively originating three and two bands. Further lowering of the symmetry determined by distortions or other effects can induce further splittings.

We investigate the simplest realization of such a mechanism, namely a
system of three bands (the minimal situation in order to have manifolds
of different degeneracy after the splitting, i.e. two degenerate bands and one lifted by the crystal field) of equal bandwidth with 4 electrons per site.
In absence of the crystal field splitting each band will be populated by 4/3 electrons. If we continuously lift one of the bands to higher energy, the electrons will gradually move to the lower levels. Therefore the density of the lifted band will decrease from 4/3 eventually reaching 1, becoming half-filled. Then, if the interaction strength is enough to localize the half-filled band, but it is smaller than the critical value for the remaining three electrons hosted by the lower two bands, we can expect an OSMT.

A first step in this direction has been taken in Ref.~\onlinecite{Dai_OSMT}, where an OSMT has been reported for two wide degenerate bands and a narrower one lifted in energy, occupied by 4 electrons per site. Unfortunately in that model both the difference in bandwidth and the lifted degeneracy mechanisms are at work and none could be singled out as the driving one.

The electrons in the three bands are coupled via a local $SU(2)$ invariant interaction. The hamiltonian reads 
\bea\label{H_int}
 H=&-&t\@ \sum_{<ij>, m \sigma}(d^\dagger_{im\sigma}d_{jm\sigma}+ {\rm h.c})
 +\sum_{i,m\sigma}\e_m d^\dagger_{im\sigma}d_{im\sigma}\nonumber \\
 \@&+\@&\# U\sum_{i,m}
 n_{im\uparrow} n_{im\downarrow}+(U' -\frac{J}{2})\@\sum_{i,m>m' } n_{im} n_{im'} \\
 \@&-\@&\# J\@\sum_{i,m>m'}\@\# \left [ 2 {\bf S}_{im}\#  \cdot {\bf S}_{im'}\!
 +(d^\dagger_{im\uparrow}d^\dagger_{im\downarrow}d_{im'\uparrow}d_{im'\downarrow}\#+h.c.)\right]\# .\nonumber
 \eea
Here $d_{i,m\s}$ is the destruction operator of an electron of spin $\s$ at site i in orbital m, and $n_{im\s}\equiv d^\+_{im\s}d_{im\s}$, $n_{im}\equiv \sum_\s d^\+_{im\s}d_{im\s}$, ${\bf S}_{im}$ is the spin operator for orbital m at site i, t is the nearest-neighbor hopping (denoted in the sum by $<>$), $\eps_m$ is the bare energy level in orbital m.
$U$ and $U'=U-2J$ are intra- and  inter-orbital repulsions and $J$ is the Hund's coupling. The densities of states of the three bands are semicircular of half-bandwidth D.

\begin{figure}[htbp]
\begin{center}
\includegraphics[width=7.5cm, height=5cm]{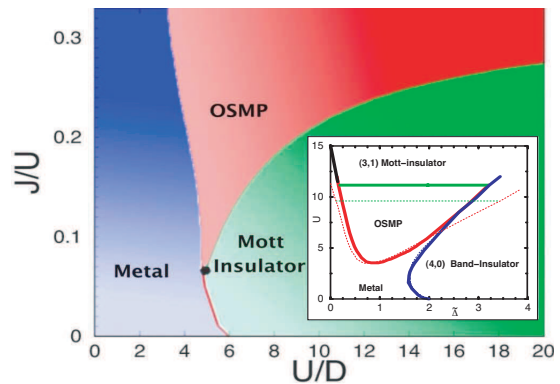}
\end{center}
\caption{(Color online)  Phase diagram for fixed populations $n_m=(1,1.5,1.5)$ (obtained by adjusting the crystal field $\D$) within Slave-spin mean-field. Inset: phase diagram for fixed total filling $n=4$ as a function of $U$ and $\D$ at $J/U=0.25$. Dashed lines: modification of this diagram under a small splitting ($\sim 0.4/D$) of the two degenerate bands.}
\label{fig:PHDiag_SS}
\end{figure}

We study this three-band model assuming that two bands have the same energy ($\eps_2=\eps_3$) and one is lifted by a crystal field splitting $\Delta\equiv \eps_1- \eps_2>0$. 
The $\Delta<0$ case, which is believed to be relevant to $Sr_{2-x}Ca_xRuO_4$ has been studied in \cite{Liebsch_SrRuO4_negDelta} and does not lead to an OSMT. Yet, it has been recently proposed that a similar mechanism to what we present here applies in $Sr_{0.2}Ca_{1.8}RuO_4$ even if $\Delta<0$ thanks to a doubling of the unit cell\cite{Neupane_OSMT}.

\begin{figure}[htbp]
\begin{center}
\includegraphics[width=7.5cm]{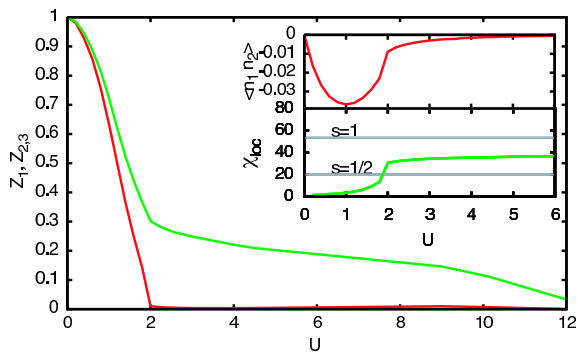}
\end{center}
\caption{(Color online)  Quasiparticle residue as a function of the interaction strength for $J/U=0.25$ and fixed populations $n=(1,1.5,1.5)$ in DMFT with $N_s=9$. Inset: interorbital correlations (top panel) and local spin susceptibility (bottom panel). The cutoff "temperature" (see text) is $\beta D=80$.}
\label{fig:Z}
\end{figure}

We use two local mean-field approximations: the faster and
computationally inexpensive slave-spin
mean-field\cite{demedici_Slave-spins,Hassan_CSSMF} (SSMF) for surveying
the phase diagram and DMFT, solved with exact diagonalization (ED), for
more accurate and  aimed calculations. In fig. \ref{fig:PHDiag_SS} we
show the SSMF phase diagram obtained adjusting $\D$ in order to always
have 1 electron in the lifted band, and 1.5 electrons in each of the
degenerate ones. Indeed an orbitally-selective Mott phase (OSMP) is
found for a large zone of the parameters $U$ and $J$. It is worth noting
that a finite Hund's coupling is needed to stabilize the OSMP, 
while for small $J$ a direct transition from a metal to a Mott insulator is found.
The indications of SSMF are confirmed by the more accurate DMFT, as shown in Fig, \ref{fig:Z}, where we plot
$Z_{\alpha}=(1-\Im\Sigma_{\alpha}(i\omega_0)/\omega_0)^{-1}$ 
($\Sigma_{\alpha}(\omega)$ 
 being the self-energy for the band $\alpha$), which measures the
 low-frequency spectral weight associated with metallic behavior. $Z_1$ for the lifted band vanishes at a critical $U$, signaling the localization of this band, while the same quantity is still finite for the two lower bands (The data are for $J/U=0.25$). We notice that ED calculations suffer from truncation effects. Analyzing these effects we find that the actual $U_c$ will be higher than what shown in the figure and we estimate the DMFT value of $U_c \simeq 2.5D$. Comparison with SSMF confirms the reliability of the latter approach, which only slightly overestimates $U_c$.

The phase diagram clearly shows that increasing $J/U$ increases the
region of the OSMP. We can gain more insight
analyzing the orbital fluctuations $\langle n_1 n_2\rangle - \langle n_1
\rangle \langle n_2\rangle$. In order to have an OSMT this quantity
should be small, signaling a decoupling of the bands which opens the way
for a different behavior between them, and the localization of the half-filled one.
As shown in the inset of Fig. \ref{fig:Z}, for $J=U=0$ the orbitals are
uncorrelated. Increasing the two quantities, $U$ initially prevails,
leading to an increased orbital correlation. 
Further increasing $U$ and $J$ makes the electrons more and more
localized. 
In this regime the effect of $J$ becomes
predominant\cite{capone:science}, and it reduces the orbital
correlations.
The role of $J$ can be understood in the atomic limit (which is reached for
very large $U$). In this limit increasing $J$ enhances the distance between the 
lowest-lying high-spin state in which the orbitals are equally
populated, and the multiplets with different orbital population.
Thus fluctuations between the two orbitals are suppressed.
The effect of $J$ is clearly stronger in the presence of the
crystal-field splitting that naturally decouples the orbitals.
The $U_c$ for the OSMT slowly approaches the single band one
($U_c=3.34D$ in SSMF) as $J/U$ increases. 
The transition for the other bands instead gets pushed at larger $U/D$,
so that the OSMP zone widens at large $J$. 

It is important, in order to assess the generality of this phase, to show that the OSMT occurs for a finite range of $\D$ and that no fine-tuning is needed. We have thus fixed only the total population to 4 electrons per site and studied the system as a function of $\D$ for several values of $U$ at fixed $J/U=0.25$. The phase diagram obtained in SSMF is plotted in in the inset of figure \ref{fig:PHDiag_SS} and shows that the region of the OSMP is quite large, and it increases with $U$. The critical $\D$ assumes reasonable values already for intermediate coupling.
We notice that this phase diagram bares resemblance to what found in Ref. \cite{Dai_OSMT} for a version of our model in which the bandwidth of the lifted band is half of the other two, suggesting  that also in that case the driving mechanism for the OSMT is the different degeneracy and not the bandwidth difference.

\begin{figure}[htbp]
\begin{center}
\includegraphics[width=8cm]{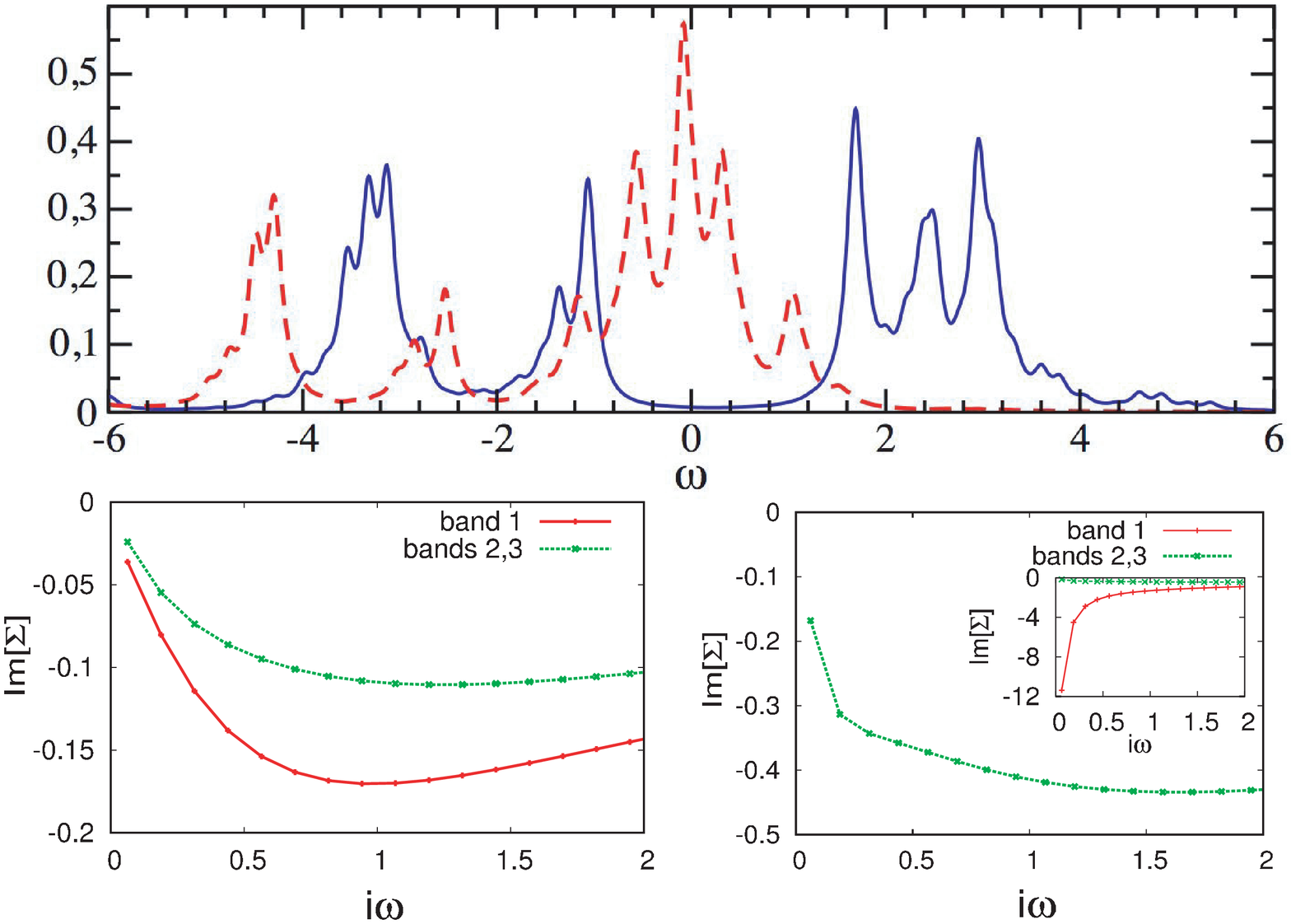}
\end{center}
\caption{(Color online)  Upper panel: local spectral densities for the lifted (solid blue) and the degenerate (dashed red) bands in the Orbital-selective Mott phase. 
(calculations in DMFT-ED with $N_s=12$ for $U/D=4.5$). Bottom panels: self-energies respectively in the metallic ($U/D=1.0$ - left) and orbitally selective ($U/D=2.4$ - right) phases.}
\label{fig:spectral}
\end{figure}

In Fig. \ref{fig:spectral} we show the spectral densities of the two bands. As previously studied\cite{demedici_Slave-spins,Ferrero_OSMT} for finite $J$ the insulating band shows a full gap, while the other two show spectral weight at the chemical potential. This gives an immediate explanation of the phase diagram in the $\D$-$U$ plane (inset of Fig.\ref{fig:Z}): there is a large range of $\D$ for which the chemical potential falls inside the gap of the localized band, leaving it insulating.

A key quantity for understanding the nature of the OSMP is the local spin susceptibility $\chi_{loc}$ (DMFT results in the second panel in the inset of Fig.~\ref{fig:Z}).
When a band gets localized $\chi_{loc}$ goes from a Pauli-like ($\sim const.$) to a Curie-like ($\sim 1/T$) behavior indicating the formation of free moments. At $T=0$ the susceptibility therefore diverges, but in the ED calculation the fictive temperature $\beta$ plays the role of a low-energy cutoff. As a result the saturation value is $\beta S(S+1)/3$. 
In a one-band Mott insulator $\chi=\beta/4$, signaling $S=1/2$\cite{georges_RMP_dmft}, while for an $N$-band half-filled Mott insulator with $J > 0$ the high-spin state $S=N/2$ is selected\cite{pruschke_Hund}, and one finds 
$\chi=\beta N/2(N/2+1)/3$.

When an OSMT occurs, we expect a two-stage saturation as a function of $U$, with the lifted band localizing before the complete Mott transition occurs. Therefore, in the OSMP $\chi_{loc}$  is expected to show a Curie component due to the magnetic moment of the localized electrons and a contribution of the itinerant electrons. The Hund's coupling between the different orbitals will actually induce a local moment in the itinerant bands. Thus the total effective moment will be larger than that of the localized component\cite{Biermann_nfl}.
The DMFT results clearly confirm these expectations, as the momentum is intermediate between the spin-1/2 of an independent single-band Mott state and spin-1, as shown in Fig.~\ref{fig:Z}. 

Another interesting property of the OSMP is the non Fermi-liquid nature of the itinerant component. The magnetic moments of the localized electrons act as scattering centers for the itinerant electrons. In the two-band model with different bandwidths, for an $SU(2)$ Hund's coupling, a logarithmic  behavior\cite{Biermann_nfl} at low $\omega$  or a power-law\cite{Werner_nfl} has been reported for the imaginary part of the self-energy.
In the present model the non Fermi-liquid behavior is confirmed. Indeed as shown in Fig. \ref{fig:spectral}, while in the metallic phase the self-energies have a linear behavior at low $\omega$, in the OSMP the metallic one shows a much faster drop.

\begin{figure}[htbp]
\begin{center}
\includegraphics[width=7cm]{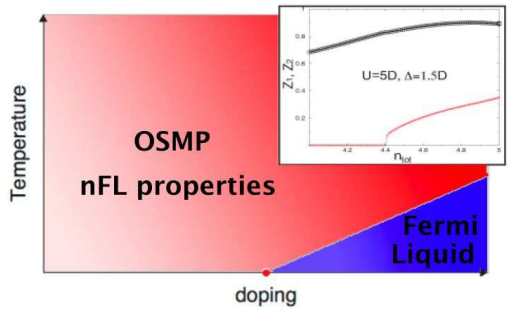}
\end{center}
\caption{(Color online)  Sketch of the general phase diagram of a doped OSMP. Inset: quasiparticle residues as a function of the filling in SSMF for $U=5$, $J/U=0.25$ and $\D=1.5$. The OSMP turns into a Fermi liquid at a finite doping.}
\label{fig:sketch}
\end{figure}

We finally investigate the stability of the OSMP by respect to doping. 
The ``standard'' Mott insulating phase in the single-band and in
degenerate multiband models is destroyed by any finite doping, while, as
mentioned in the introduction, in Ref. \cite{Werner_Hund} an OSMT is
found doping a half-filled two-band Mott insulator with a Hund's
coupling if the two levels are shifted. In our  model a similar
mechanism applies: in the OSMP the chemical potential can move within
the gap of the localized band (See Fig.~\ref{fig:spectral}) without
doping it, and at the same time dope the gapless metallic bands without
spoiling the OSMP. Then at some critical doping we expect the chemical potential to hit the Hubbard band of the localized electrons, leading to a doping also of this band, and to its metallic behavior.
Indeed numerical results confirm this scenario that bares substantial analogies with previous calculations on doped OSMP's\cite{Koga_OSMT,Ruegg_OSMT}. As we see in the inset of Fig. \ref{fig:sketch} for a set of parameters not far from the border of the phase, the OSMP is quite robust with respect to doping, then eventually at a \emph{finite critical doping}  the OSMP leaves the place to a normal Fermi liquid.
It is interesting to notice that the general phase diagram of a doped OSMP is reminiscent of the behavior of the iron-pnictide superconductors (and also of cuprates). Doping leads from a ``strange phase'' in which metallic and insulating properties coexist, to a more regular Fermi-liquid. 

From a general perspective, the electronic configuration of the pnictides\cite{Haule_FeAs} is ideal for our mechanism to apply: coulomb correlations and Hund's coupling are sizeable and the population of 6 electrons in 5 bands is even more favorable than the one we studied. Here the starting point in the absence of splittings is $n=6/5$ for each band, suggesting that a smaller crystal-field splitting may be sufficient to make the highest band half-filled, and therefore localized (or quasi-localized).

In summary we have outlined a new mechanism for the OSMT: a crystal field can split a multiband system in manifolds of different degeneracy, in which correlated electrons have different kinetic energy despite the equal bandwidth. If  the crystal field and the Hund's coupling are large enough, the suppression of the orbital fluctuations decouples the manifolds and allows a selective localization of the lifted band at intermediate couplings. Importantly, the result occurs in a wide range of model parameters, suggesting that this mechanism can be realized in several systems\cite{Nekrasov_LiV2O4}.
We have also checked (see Fig. \ref{fig:PHDiag_SS}) that the OSMP is solid with respect to further degeneracy lifting of  the lower bands, at least for small to intermediate splittings. Even if all the electrons come from bands of the same bare dispersion, the OSMP is a highly non conventional metal resulting from the selective localization:
it has a non Fermi-liquid behavior and a Curie-Weiss magnetic response, and it is generally unstable to a normal Fermi-liquid phase at a finite doping.

We acknowledge valuable discussions with A. Camjayi, H. Ding, K. Haule, G. Kotliar, A.-M.S Tremblay and Z. Fang, and support from the Center for Materials Theory of Rutgers University, NSERC (Canada), Italian MIUR PRIN 2007,  NSF of China and the 973 program of China (No.2007CB925000).
Slave spin computations were performed on the Dell cluster of the RQCHP.



\begin{thebibliography}{27}
\bibitem{georges_RMP_dmft} A.~Georges {\sl et al.}, Rev. Mod. Phys. {\bf 68}, (1996).

\bibitem{limelette_v2o3_science}P.~Limelette {\sl et al.}, Science {\bf 302}, 89 (2003).

\bibitem{Anisimov_OSMT} V.~Anisimov {\sl et al.}, {\bf 25}, 191 (2002).

\bibitem{Koga_OSMT} A.~Koga {\sl et al.}, Phys. Rev. Lett. {\bf 92}, 216402 (2004).

\bibitem{demedici_Slave-spins} L.~de' Medici, A.~Georges and S.~Biermann, Phys. Rev. B {\bf 72}, 205124 (2005).

\bibitem{Ferrero_OSMT} M.~Ferrero {\sl et al.}, Phys. Rev. B {\bf 72}, 205126 (2005).

\bibitem{Liebsch_OSMT_3} A.~Liebsch, Phys. Rev. Lett. {\bf 95}, 116402 (2005).

\bibitem{Biermann_nfl} S.~Biermann, L.~de' Medici and A.~Georges, Phys. Rev. Lett. {\bf 95}, 206401 (2005).

\bibitem{Dai_OSMT} X.~Dai, G.~Kotliar and Z.~Fang,  arXiv:cond-mat/0611075v1.

\bibitem{Liebsch_SrRuO4_negDelta}A.~Liebsch and H.~Ishida, Phys. Rev. Lett. {\bf 98}, 216403 (2007).

\bibitem{Neupane_OSMT} M.~Neupane {\sl et al.}, arXiv:0808.0346v1.

\bibitem{Balicas_fsfceCSRO} L.~Balicas {\sl et al.}, Phys. Rev. Lett. {\bf 95}, 196407 (2005).

\bibitem{Wang_Yang_ARPES_CaSrRuO4} S.-C.~Wang {\sl et al.}, Phys. Rev. Lett. {\bf 93}, 177007 (2004).

\bibitem{kamihara} Y.~Kamihara {\sl et al.}, J. Am Chem. Soc. {\bf 128}, 10012 (2006);
{\it ibid.} {\bf 130}, 3296 (2008)

\bibitem{Shorikov} A.~O. Shorikov {\sl et al.},  arXiv:0804.3283v2.

\bibitem{Venturini_MIT_cuprates} F.~Venturini {\sl et al.}, Phys. Rev. Lett. {\bf 89}, 107003 (2002).

\bibitem{Wu_FeAs_OSMT} J.~Wu, P.~Phillips and A.~Castro~Neto, arXiv:0805.2167v1.

\bibitem{Werner_Hund} P.~Werner and A.~Millis, Phys. Rev. Lett. {\bf 99}, 126405 (2007).

\bibitem{gunnarsson_fullerenes}
O.~Gunnarsson, E.~Koch and R.~Martin, Phys. Rev. B {\bf 54}, R11026 (1996).

\bibitem{florens_multiorb} S.~Florens {\sl et al.}, Phys. Rev. B {\bf 66}, 205102 (2002).

\bibitem{Lu_gutz_multiorb} P.~Lu, Phys. Rev. B {\bf 49}, 5687 (1994).


\bibitem{rozenberg_multiorb} M.~J.~Rozenberg, Phys. Rev. B {\bf 55}, R4855 (1997).

\bibitem{Hassan_CSSMF} S.R.Hassan and L.~de'~Medici, arXiv:0805.3550v1.

\bibitem{capone:science} M.~Capone, M.~Fabrizio,  C.~Castellani and E.~Tosatti, Science {\bf 296}, 2364 (2002).

\bibitem{pruschke_Hund} T.Pruschke and R.~Bulla, Eur. Phys. J. B {\bf 44}, 217 (2005).

\bibitem{Werner_nfl} P.~Werner and A.~Millis, Phys. Rev. B {\bf 74}, 155107 (2006).

\bibitem{Ruegg_OSMT} A.~Ruegg {\sl et al.}, Eur. Phys. J. B {\bf 48}, 55 (2005).

\bibitem{Haule_FeAs} K.~Haule and G.~Kotliar, arXiv:0805.0722v1.

\bibitem{Nekrasov_LiV2O4} I.~A.~Nekrasov {\sl et al.}, Phys. Rev. B {\bf 67}, 085111 (2003).

\end{thebibliography}
\end{document}